\documentstyle[preprint,tighten,aps,psfig]{revtex}
\begin{document}
\title{ $SU(2)_L\times SU(2)_R$ and 
$U(1)_A$ restorations high in the hadron spectrum
and what  it tells
 us about.}
\author{ L. Ya. Glozman}
\address{  Institute for Theoretical
Physics, University of Graz, Universit\"atsplatz 5, A-8010
Graz, Austria\footnote{e-mail: leonid.glozman@uni-graz.at}}
\maketitle

\begin{abstract} 
Recent data for highly excited mesons suggest
that not only the chiral $SU(2)_L\times SU(2)_R$ symmetry of QCD is
restored high in the spectrum but also the $ U(1)_A$
symmetry. This 
means that it is not a confining interaction in QCD which
triggers the spontaneous breaking of chiral symmetry.
 The restoration of the 
$U(2)_L\times U(2)_R$
symmetry  of the QCD Lagrangian  implies the appearance
of  multiplets of this  group high in the hadron spectra.
Such type of multiplets  is naturally explained within the  string
picture of confinement.
It also  supports the scenario that the  $U(1)_A$
breaking is related to instantons and not to the gluonic interaction
responsible for confinement.
\end{abstract}

\bigskip
\bigskip

\section{Introduction}

It has recently been suggested that high in the hadron spectra
the spontaneously broken chiral symmetry of the QCD Lagrangian
is effectively restored \cite{G}. This phenomenon can be
understood in very general terms from the well established
concepts of quark-hadron duality, the validity of the operator
product expansion in QCD at large space-like momenta, and
the validity of the dispersion relation for the two-point
correlator, which connects the space-like and time-like
regions \cite{CG1,CG2}. The phenomenological manifestation
of the effective chiral symmetry restoration is that the
high-lying hadrons in the ($u,d$) quark sector must fall into
multiplets of the parity-chiral group; they manifest themselves
as parity doublets or higher multiplets containing degenerate
states of opposite parity \cite{CG1,CG2}. This phenomenon does
not mean that the spontaneous breaking of chiral symmetry in
the QCD vacuum disappears, but rather that it
becomes irrelevant once we are sufficiently
high in the spectrum. While the chiral symmetry breaking
condensates are crucially important for the physics of the
low-lying states, the physics of the high-lying hadrons is
such as if there were no chiral symmetry breaking in the vacuum.
A very natural picture for the highly excited hadrons then is 
that they represent  rotating strings (with the color-electric
field in the string) with practically massless bare
 quarks of definite chirality at the ends
of the string and these valence quarks
are combined into the parity-chiral multiplets \cite{G2}.\\

The general phenomenon of chiral  $ SU(2)_L\times SU(2)_R$
symmetry restoration high in the spectrum as well as the physical
picture of excited hadrons as strings is also well compatible
with the restoration of the higher symmetry  $ U(2)_L \times U(2)_R$.
For the latter it is necessary that not only the chiral symmetry
is restored but also the $U(1)_A$ symmetry. For the restoration
of the $U(1)_A$ symmetry  both the explicit $U(1)_A$
breaking through the axial anomaly and the 
spontaneous $ U(1)_V \times U(1)_A  \rightarrow U(1)_V$ breaking
must become unimportant \cite{CG1}. While the effective
chiral symmetry restoration
high in the spectrum  implies that the spontaneous 
$ U(1)_V \times U(1)_A$ breaking must also become irrelevant
(because both
are broken by the same quark condensates in the QCD vacuum),
it does by no means cause  the effects of the explicit $U(1)_A$
breaking to disappear high in the spectra. The latter  could only happen
 if the gluonic interactions that couple to the flavor-singlet
current via the axial anomaly became unimportant due to some reason.\\

The main purpose of this Letter is to show that  recent data on
 highly excited mesons obtained from the partial wave analysis of
proton-antiproton annihilation at LEAR \cite{BUGG1,BUGG2}  indicate
that not only  the chiral symmetry but also
the full $ U(2)_L \times U(2)_R$ symmetry of the QCD Lagrangian get
restored.\\

The second purpose is to argue which possible picture could be
behind both the chiral and $U(1)_A$  symmetries restoration high in the
hadron spectra.
Historically the $U(1)_A$ problem began with the observation
that the three-flavor singlet state $\eta \prime$ is too heavy to be
considered as a ninth (pseudo)Goldstone boson, which would be
required if the pattern of spontaneous symmetry breaking
were $ U(3)_L\times U(3)_R \rightarrow U(3)_V$ \cite{WEIN}.
While the Gell-Mann - Oakes - Renner relations (GOR) \cite{GOR}
are very successful for the octet states, $\pi, K, \eta$,
and the masses of these particles do vanish with the bare
quark masses, this is apparently not the case for the
flavor-singlet state $\eta \prime$ in which case the GOR fails.\\

't Hooft has suggested a very elegant solution for this
problem \cite{HOOFT}. He realized that the coupling of quarks
to the localized topological solutions of the pure gluonic
field (instantons) \cite{BPST} via the axial anomaly produces
an effective interaction between quarks which is still
$ SU(N_f)_L\times SU(N_f)_R$ invariant but breaks explicitly
the $U(1)_A$ symmetry and is repulsive
in the flavor-singlet pseudoscalar channel. Hence the instantons
break the $ U(N_f)_L\times U(N_f)_R$ symmetry
of the QCD Lagrangian in the chiral limit to the
lower symmetry  $ SU(N_f)_L\times SU(N_f)_R\times U(1)_V$ and thus
the lowest flavor-singlet pseudoscalar meson cannot be considered
any longer  a (pseudo)Goldstone boson.\\

Later on Witten \cite{WITTEN} and  Veneziano
\cite{VENEZIANO}  argued that the solution of
the $U(1)_A$ problem should not necessarily come from
instantons 
but rather from some other type of gluonic
configurations in QCD (e.g. the ones related to confinement),
 which also couple to the flavor-singlet quark state
via the axial anomaly. This resulted in the famous
Witten-Veneziano formula which relates the flavor-singlet
mass to the topological susceptibility in pure gauge
theory and  also shows that in the large $N_c$ limit,
where the axial anomaly vanishes, the $ U(N_f)_L\times U(N_f)_R$
symmetry is restored and the flavor-singlet pseudoscalar
state becomes the (pseudo)Goldstone boson 
of the spontaneously broken $ U(N_f)_L\times U(N_f)_R$
symmetry like the octet states.\\

A related problem  may be the origin
of chiral symmetry breaking in QCD. At approximately
the same time Caldi as well as Callan, Dashen and Gross 
and others \cite{CDG}
have suggested that instantons under some conditions
could also provide the
spontaneous chiral symmetry breaking in the QCD vacuum. The 
mechanism of chiral symmetry breaking is rather obvious --
the 't Hooft interaction contains  the Nambu and
Jona-Lasinio \cite{NJL} 4-fermion vertex and hence can trigger
the chiral symmetry breaking. The phenomenological arguments
for the instanton liquid  structure of the vacuum 
as well as the typical size and separation between instantons
have been given by Shuryak \cite{SHURYAK}. Diakonov and Petrov
have derived a microscopical theory of chiral symmetry breaking
by instantons \cite{DP}. Since then  much work has been done
by different groups and the present state of the art is
summarized in review \cite{SHS}. It should be stressed that
there are both strong theoretical as well as lattice arguments
that the instanton medium alone cannot generate the confinement mechanism
in QCD and hence some additional gluonic configurations are
required in the QCD vacuum in order to explain  confinement,
chiral symmetry breaking, and $U(1)_A$  symmetry breaking, if instantons are
indeed important for the $U(1)_A$ and chiral symmetry breaking.
So it is evident that the instanton liquid structure of the
QCD vacuum cannot be a complete picture of the vacuum. Indeed
the instantons  provide only a
very small part of the full action of QCD.\\

These issues, and in particular, to which extent instantons
contribute (or not) to the structure of the QCD vacuum,
to chiral symmetry breaking, and to the structure of the
low-lying hadrons, are  vividly discussed topics in the lattice
community. Starting first with the technique of  cooling
of the gluonic part of the action \cite{NEG}, these studies have 
moved to the point to which extent fermions see the instantons
(via the would-be-zero modes). There is still a controversy
as one of the lattice
groups  argues against the instantons \cite{LATTICE1}. However,
their results have  been questioned and cross-checked 
by the other groups, which see evidence in favour of instantons
 \cite{LATTICE2}. 
For example, the results \cite{LATTICE3} show in particular a direct 
correlation
of the would-be-zero modes of fermions and the self-dual or
anti-self-dual lumps of gluonic field (see the second paper in 
the reference above). While these studies do evidence
that instantons are very important for chiral symmetry breaking,
it is still unclear to which extent they can explain it, whether
it is the effect of instantons only or a combined effect of instantons and
something else.\\

We want to present an empirical
argument in favour of the point of view
that it is instantons that are responsible for $U(1)_A$
breaking in QCD.   
The experimental data suggest
that high in the hadron spectrum both the 
$ SU(2)_L\times SU(2)_R$ and $  U(1)_A$
symmetries are approximately restored. 
Since the physics of the highly-excited states
is most probably due to the confinement in QCD, one can
conclude (at least preliminary) that it is not a confining
interaction in QCD which is responsible for both the
$ SU(2)_L\times SU(2)_R$ and $  U(1)_A$
breakings. While it  suggests that instantons are indeed important
for the $U(1)_A$ breaking, these data cannot shed any light on whether
only instantons or instantons and something else
provide the  chiral symmetry breaking in the QCD vacuum.\\

\section{Empirical pattern of the
$ SU(2)_L\times SU(2)_R \times U(1)_V\times U(1)_A$ symmetry breaking}

In this Letter we limit ourselves to the two-flavor version
of QCD. There are two reasons for doing this. First of all,
the $u$ and $d$ quark masses are very small as compared
to $\Lambda_{QCD}$ and the typical hadronic scale of 1 GeV. Thus
the chiral $ SU(2)_L\times SU(2)_R$ and more generally the
$ U(2)_L\times U(2)_R$ symmetries of the QCD Lagrangian are
nearly perfect. This is not the case if the $s$ quark
is included, and {\it a-priori} it is not clear whether one should
regard this quark as light or "heavy". The second reason
is a practical one -- there are good new data on highly excited
$u,d$ mesons, but such data are still missing for the strange
mesons. Certainly it would be very interesting and important
to extend the analysis of the present paper to  the 
$ U(3)_L\times U(3)_R$
case. We hope that the present results will stimulate the 
experimental and theoretical work in this direction.\\

 The lowest pseudoscalar and scalar mesons give
an idea of how strongly the $ SU(2)_L\times SU(2)_R$ and
$ U(1)_V\times U(1)_A$ symmetries are broken. In QCD the meson
masses  are extracted
from the two-point correlator,

\begin{equation}
\langle 0 | T \left \{ j_\alpha(x) j_\alpha(0)\right \} | 0 \rangle,
\label{corr}
\end{equation}

\noindent
where $\alpha$ specifies the set of quantum numbers of
the current (interpolating field), $j_\alpha(x)$; it
coincides with the set of quantum numbers of the meson of interest.
All the information about the hadron spectrum is encoded in the
complicated structure of the QCD vacuum and the physical hadrons
with the quantum numbers $\alpha$ represent a response of the
vacuum to the external probe $j_\alpha(x)$. For the  pseudoscalar
and scalar mesons $\pi, f_0, a_0$ and $\bar \eta$ \footnote{
The $\bar \eta$
represents the singlet state in two-flavor QCD which is
analogous to the flavor-singlet state $\eta \prime$ in  three-flavor QCD;
its mass can be approximately extracted from the masses of physical
$\eta$ and $\eta \prime$ mesons by unmixing the $(u\bar u + d\bar d)/\sqrt 2$
and $s\bar s$ components - see Appendix.} the interpolating fields
are given as

\begin{equation}
 j_\pi(x)  = \bar q(x) \frac{\vec \tau}{2} \imath \gamma_5 q(x),
\label{pi}
\end{equation}

\begin{equation}
 j_{f_0}(x)  = \frac{1}{2}\bar q(x)  q(x),
\label{f0}
\end{equation}

\begin{equation}
 j_{\bar \eta}(x)  = \frac{1}{2}\bar q(x)  \imath \gamma_5 q(x),
\label{pi}
\end{equation}

\begin{equation}
 j_{a_0}(x)  = \bar q(x) \frac{\vec \tau}{2}  q(x).
\label{pi}
\end{equation}

\noindent
These four currents belong to the
irreducible representation
of the 
$U(2)_L\times U(2)_R = SU(2)_L\times SU(2)_R \times U(1)_V\times U(1)_A$
group. It is instructive to see how these currents transform
under different subgroups of the group above.
The irreducible representations of $SU(2)_L\times SU(2)_R$
can be labeled as $(I_L, I_R)$ with $I_L$ and $I_R$
being the isospins of the left and right subgroups. However,
generally the states that belong to the given irreducible representation
of the chiral group cannot be ascribed a definite parity
because under parity transformation the left-handed quarks
transform into the right-handed ones (and vice versa). Therefore under
a parity operation the irreducible representation $(I_L, I_R)$ transforms
into $(I_R, I_L)$. Hence, in general, the state (or current)
 of definite parity can
be constructed as a direct sum of two irreducible representations
$(I_L, I_R) \oplus (I_R, I_L)$, which is an irreducible
representation of the parity-chiral group \cite{CG2}.\\

 The $SU(2)_L\times SU(2)_R$ transformations consist of vectorial
 and axial transformations in the isospin space. The axial
transformations mix the currents of opposite parity:

\begin{equation}
 j_\pi(x)  \leftrightarrow j_{f_0}(x) 
\label{pif0}
\end{equation} 

\noindent
as well as

\begin{equation}
 j_{a_0}(x)  \leftrightarrow j_{\bar \eta}(x).
\label{a0eta}
\end{equation} 

\noindent
Each pair of currents belongs to the $(1/2,1/2)$
representation of the parity-chiral group, which contains both
$I=0$ as well as $I=1$ states.\\

The $U(1)_A$ transformation mixes the currents
of the same isospin but opposite parity:

\begin{equation}
 j_\pi(x)  \leftrightarrow j_{a_0}(x) 
\label{pia0}
\end{equation} 

\noindent
as well as

\begin{equation}
 j_{f_0}(x)  \leftrightarrow j_{\bar \eta}(x).
\label{f0eta}
\end{equation}

\noindent
All four currents together belong to the irreducible representation
$(1/2,1/2) \oplus (1/2,1/2)$ of the $U(2)_L\times U(2)_R $ group.\\

If the vacuum were invariant with respect to $U(2)_L\times U(2)_R $
transformations, then all four mesons, $\pi,f_0,a_0$ and $\bar \eta$
would be degenerate (as well as all their excited states). Once
the $U(1)_A$ symmetry is broken explicitly through
the axial anomaly, but the chiral $SU(2)_L\times SU(2)_R $ 
symmetry is still
intact in the vacuum, then the spectrum would consist of
degenerate $(\pi, f_0)$ and $(a_0, \bar \eta)$ pairs. If
in addition the chiral  $SU(2)_L\times SU(2)_R $ symmetry is
spontaneously broken 
in the vacuum, the degeneracy is also lifted in  the pairs
above and the pion becomes a (pseudo)Goldstone boson. Indeed,
the masses of the lowest mesons  are \cite{PDG}

 $$ m_\pi \simeq 140 MeV, ~m_{f_0} \simeq 400 - 1200 MeV,~
m_{a_0} \simeq 985 MeV ,~ m_{\bar \eta} \simeq 782 MeV. $$

This immediately tells that both $SU(2)_L\times SU(2)_R $ and
$U(1)_V \times U(1)_A$ are broken in the QCD vacuum
 to $SU(2)_I$ and $U(1)_V$, respectively.\footnote
{ In the chiral symmetry broken regime the use of effective
degrees of freedom in the low-lying hadrons
is certainly fruitful. The chiral symmetry
breaking implies that practically massless quarks acquire
a quasiparticle (dynamical or constituent) mass through
their coupling to the quark condensates of the vacuum. How it
happens is well seen from the schematical Nambu and Jona-Lasinio
model. 
Pions are also well understood from 
the Nambu and Jona-Lasinio picture of chiral symmetry breaking
and the formation of the lowest excitation over the vacuum
is  analogous to the Anderson mode in superconductors. In this picture
the pion is a relativistic
bound state of two quasi-particles $Q \bar Q$.
The quasiparticle $Q$ itself is a result of chiral symmetry
breaking in the vacuum.
 The "residual" 
attraction of these quasiparticles in the isovector-pseudoscalar
channel is unambiguously fixed by  chiral symmetry and once it
 is taken into account within
the Bethe-Salpeter approach it necessarily leads to the zero mass
of pions in the chiral limit. The pion is a highly collective
mode, but not a simple $q\bar q$ excitation, because the quasiparticle
$Q$ itself is a highly collective coherent excitation of  bare quarks
and antiquarks.}\\

There are a few possible scenarios that are consistent with
the given pattern.

(i) both the $SU(2)_L\times SU(2)_R $ and $U(1)_V \times U(1)_A$
breakings come from the particular gluodynamics in QCD that is responsible for
confinement;

(ii) the $SU(2)_L\times SU(2)_R $ breaking is due to the
gluodynamics that is responsible for confinement and the
$U(1)_V \times U(1)_A$ breaking comes from instantons;

(iii) the $U(1)_V \times U(1)_A$ breaking is due to the
confinement while the $SU(2)_L\times SU(2)_R $ breaking is
from other sources;

(iv) the $U(1)_V \times U(1)_A$ breaking is provided by instantons
while the $SU(2)_L\times SU(2)_R $ breaking is related to
instantons alone or to a combination of instantons
and some other possible gluonic interactions that
are not related directly to confinement.\\

In the following we will show that the scenarios (i), (ii) and (iii)
are very unlikely in view of the new empirical data on
highly excited mesons.

\section{What do we learn from the highly excited hadrons?}

Systematic data on highly excited mesons are still missing
in the PDG tables. We will use the recent
results of the partial wave analysis of mesonic resonances
obtained in $p \bar p$ annihilation at LEAR \cite{BUGG1,BUGG2}.
For the scalar and pseudoscalar mesons in the mass range
from 1.8 GeV to 2.4 GeV the corresponding 
results are summarized in  the Table below.
We note that the $f_0$ state at $2102 \pm 13$  MeV is {\it not}
considered by the authors as  a $q\bar q$ state (but rather as a
candidate for glueball) because of its very unusual decay
properties and very large mixing angle. This is in contrast to all
other $f_0$ mesons in the Table, for which  the mixing angles
are small. Therefore these mesons are regarded  as 
predominantly $u,d = n$ states.
 Hence, in the following we will
exclude the $f_0$ state at $2102 \pm 13$  from our analysis which
applies only to $n\bar n$ states.\\

\begin{center}
\begin{tabular}{|llllll|} \hline
Meson & ~I~ & $~J^P~$ & Mass (MeV) & Width (MeV) & Reference\\ \hline
$f_0$ & ~0~ & $~0^+~$  & $1770 \pm 12$ &  $220 \pm 40$ & \cite{BUGG0}\\
$f_0$ & ~0~ & $~0^+~$  & $2040 \pm 38 $ &  $405 \pm 40$ & \cite{BUGG1} \\
$f_0$ & ~0~ & $~0^+~$  & $2102 \pm 13$  &  $211 \pm 29$ & \cite{BUGG1} \\
$f_0$ & ~0~ & $~0^+~$  & $2337 \pm 14$  &  $217 \pm 33$ & \cite{BUGG1} \\
$\eta$ & ~0~ & $~0^-~$  & $2010^{+35}_{-60}$   &  $270 \pm 60$ & \cite{BUGG1}\\
$\eta$ & ~0~ & $~0^-~$  & $2285 \pm 20$   &  $325 \pm 30$ & \cite{BUGG1} \\
$\pi$ & ~1~ & $~0^-~$  & $1801 \pm 13$   &  $210 \pm 15$ & \cite{PDG} \\
$\pi$ & ~1~ & $~0^-~$  & $2070 \pm 35$   &  $310^{+100}_{-50}$ & \cite{BUGG2}\\
$\pi$ & ~1~ & $~0^-~$  & $2360 \pm 25$   &  $300^{+100}_{-50}$ & \cite{BUGG2}\\
$a_0$ & ~1~ & $~0^+~$  & $2025 \pm ?$   &  $320 \pm ?$ & \cite{BUGG2}\\
 \hline
\end{tabular}
\end{center}

The prominent feature of the data is an approximate 
degeneracy of the three highest states in the pion spectrum with
the three highest states in the $f_0$ spectrum:

\begin{equation}
\pi(1801 \pm 13) - f_0(1770 \pm 12),
\end{equation}

\begin{equation}
\pi(2070 \pm 35) - f_0(2040 \pm 38 ),
\end{equation}

\begin{equation}
\pi(2360 \pm 25) - f_0(2337 \pm 14).
\end{equation}

 This can be considered as  a manifestation
 of chiral symmetry restoration high in the spectra. The approximate
 degeneracy of these physical states indicates that the
 chiral $SU(2)_L \times SU(2)_R$ transformation properties
 of the corresponding currents (see section 2) are not violated
 by the vacuum. This means that the chiral symmetry breaking of the vacuum becomes
 irrelevant for the high-lying states and the physical states 
 above form approximately the  chiral pairs
 in the $(1/2,1/2)$ representation of the chiral group.
The physics of the highly excited hadrons is such as if there
were no chiral symmetry breaking in the vacuum.\\
\\
 
 A similar behaviour is observed from a comparison of the $a_0$ and
 $\eta$ masses high in the spectra:

 \begin{equation} 
 a_0(2025 \pm ?) - \eta(2010^{+35}_{-60}). 
\end{equation}

The authors of ref. \cite{BUGG2} are confident of the existence
of $a_0(2025)$, however it is  difficult to extract the
error bars for its mass from the existing data. 
Some of the missing states with these quantum numbers
 are still to be discovered; technically the identification
 of the  $a_0$ and $\eta$ resonances is a rather difficult
 task. \\

As  was stressed before \cite{CG1,CG2}, the chiral 
symmetry restoration high in hadron spectra does not
mean that the chiral symmetry breaking in the QCD
vacuum disappears, but rather that the chiral asymmetry of the
vacuum becomes irrelevant once we are sufficiently high in the
spectra. While the quark condensates of the QCD vacuum 
are crucially important for the physics of low-lying states and "remove"
the axial part of the chiral symmetry, thereby preventing a
parity doubling low in the hadron spectra, their role high
in the spectrum becomes progressively less important and 
eventually the
chiral symmetry is  restored.\\

It is quite natural to assume that the physics of the highly
excited hadrons is due to confinement in QCD. If so, it follows
that the confining gluodynamics is still important. On  the
other hand the
chiral symmetry breaking effects in the vacuum become irrelevant.
Then the scenarios (i) and (ii) are ruled out.\\

A very natural physical picture for the highly
excited states is that these hadrons  
are  relativistic strings (with the color-electric
field in the string) with practically massless quarks
at the ends;  these massless quarks are combined into
parity-chiral multiplets \cite{G2}.
The string picture is compatible with the chiral symmetry
restoration because there always exists a solution for the
right-handed and left-handed quarks at the end of the string
with exactly the same energy and total angular momentum.
Since the nonperturbative field in the string is pure electric
and the electric field is "flavor-blind",
the string dynamics itself is not sensitive to the specific
flavor of a light quark once the chiral limit is taken.
 This picture 
explains the empirical 
parity-doubling because for every intrinsic quantum state
of the string there necessarily appears parity doubling
of the states with the same total angular momentum of hadron.
Hence the string picture is compatible not only  with the 
$SU(2)_L \times SU(2)_R$ restoration, but more generally
with the $U(2)_L \times U(2)_R$ one.\\

This picture should be contrasted with the nonrelativistic
or (semi)relativistic {\it potential} description of hadrons.
Within the potential description the parity of the state is
unambiguously prescribed by the relative orbital angular
momentum $L$ of  quarks. For example, 
 all the  states  on the radial pion Regge trajectory are 
$^1S_0$  $q \bar q$ states, while the members of the $f_0$ trajectory
are the $^3P_0$ states. Clearly, such a picture cannot
explain the {\it systematical} parity doubling as it would require
that the stronger centrifugial repulsion in the case of $^3P_0$
mesons (as compared to the $^1S_0$ ones) as well as the strong and attractive
 spin-spin force in the case of $^1S_0$ states (as compared to
 the weak  spin-spin force in the $^3P_0$ channel)
 must systematically lead to an approximate degeneracy for all 
 radial states. This is very improbable. 
  The potential
picture also implies  strong spin-orbit interactions
between quarks while the spin-orbit splittings are absent
or very small for excited mesons and baryons in the $u,d$ sector.
The strong spin-orbit interactions inevitably follow from the 
Thomas precession (once the confinement is described through a scalar
confining potential)\footnote{
Note also that a scalar potential explicitly breaks 
the chiral symmetry
in contradiction to the requirement that the chiral symmetry must
be restored high in the spectra.}, 
and this very strong spin-orbit force must be practically
exactly compensated by  other strong spin-orbit force from the
one-gluon-exchange interaction in this picture.
In principle such a cancellation could be provided by   tuning  
the parameters for some specific (sub)families of mesons. However,
in this case the spin-orbit forces become very strong for  other
(sub)families.
In contrast, in the string
picture  there is no spin-orbit force at all once the
chiral symmetry is restored \cite{G2}. That the potential description 
fails high in the spectra also follows from a comparison
of the prediction
of, e.g., ref. \cite{GI} with the recent experimental data: the potential
picture simply does not predict very many states in the region of
2 GeV. For example, while the tuning of  
parameters of the model provides an accurate description of
the three lowest states in the pion spectrum, it does not predict at
all the existence of $\pi(2070)$ and $\pi(2360)$; the forth and the
fifth radial states of the pion do not appear in this picture
up to 2.4 GeV (which means that they are predicted to be at least
$\sim 0.5$ GeV heavier than in reality). A similar situation occurs
also in other channels. The failure of the potential description is
inherently related to the fact that it cannot incorporate
chiral symmetry restoration high in the spectra.\\

The nonrelativistic or (semi)relativistic potential
picture {\it is} justified, however,
once the current quarks are heavy and move slowly
(e.g. like in charmonium and bottomonium) or  the (semi)relativistic
description can be still justified to some extent
once the proper effective
degree of freedom is a rather heavy quasiparticle but not a bare
quark
(constituent quark in the low-lying nucleons and deltas  is
not yet ultrarelativistic). 
Here the situation
is similar to  atomic physics or to the physics of positronium.
For heavy fermions the relativistic effects represent only  small 
$v^2/c^2$ corrections to the nonrelativistic picture. However,
once the quarks are ultrarelativistic, it is not justified at
all.
As a manifestation, the potential picture requires
the $f_0$ mesons to be $P$ states of quarks,  contrary
to the $S$ states of quarks in pions. On the other hand the string 
picture attributes both $\pi$ and $f_0$ states (in pairs) 
to {\it the same} intrinsic quantum state of the string with
the {\it same} angular momentum  \cite{G2}.
The opposite parity of these high-lying mesons is provided
by different right-left configurations of the quarks at the
ends of the string.\\

Upon examining the experimental data more carefully one notices
not only a degeneracy in the chiral pairs, but also an approximate 
degeneracy
in $ U(1)_A $ pairs $(\pi, a_0)$ and $(f_0, \eta)$
(in those cases where the states are established).
If so, one can preliminary conclude that not only the
chiral $SU(2)_L\times SU(2)_R $ symmetry is restored high in 
the spectra, but the whole $U(2)_L\times U(2)_R $ symmetry of
the QCD Lagrangian. Then the approximate $(1/2,1/2) \oplus (1/2,1/2)$
multiplets of this group are given by:

\begin{equation}
 \pi(1801 \pm 13) - f_0(1770 \pm 12) - a_0 (?) -
 \eta(?);
\label{quartet1}
\end{equation}

\begin{equation}
 \pi(2070 \pm 35) - f_0(2040 \pm 40) - a_0 (2025 \pm?) - 
 \eta(2010^{+35}_{-60});
\label{quartet2}
\end{equation}

\begin{equation}
 \pi(2360 \pm 25) - f_0(2337 \pm 14) - a_0(?) -  
 \eta(2285 \pm 20).
\label{quartet3}
\end{equation}   

\noindent

 This
preliminary conclusion would be strongly supported by a discovery
of the missing $a_0$ meson in the mass region around 2.3 GeV
as well as by the missing $a_0$ and $\eta$ mesons in the 1.8 GeV region.
This would also rule out the scenario (iii)
and only the scenario (iv) would be viable.\\

We have to stress, that the $U(1)_A $ restoration
high in the spectra does not mean that the axial anomaly
of QCD vanishes, but rather that the specific gluodynamics
(e.g. instantons) that are related to the anomaly 
become unimportant there.\\

It should also be emphasized 
that the only restoration of  $U(1)_V \times U(1)_A $ symmetry
(without the $SU(2)_L\times SU(2)_R $) is impossible. This was discussed
in ref. \cite{CG1}. The reason is that even if the effects
of the explicit $ U(1)_A $ symmetry breaking via the axial anomaly
vanish, the $U(1)_V \times U(1)_A $ would  still be spontaneously
broken once the $SU(2)_L\times SU(2)_R $ were spontaneously broken.
This is because the same quark condensates in the QCD vacuum
that break $SU(2)_L\times SU(2)_R $ do also break $U(1)_V \times U(1)_A $.\\

\section{How do the instantons do the job?}

The present analysis suggests that   indeed instantons
 cause the $U(1)_A$ breaking.
Then it is instructive to outline the {\it possible}
scheme of how this happens and why instantons
are not important high in the spectra.\\

The instanton-induced interaction between quarks 
in the two-flavor
case  is given as

\begin{equation}
H_{int} \sim -G \left\{[\bar q(x) q(x)]^2 +
[\bar q(x) \vec \tau \imath \gamma_5 q(x)]^2 -
[\bar q(x) \vec \tau  q(x)]^2  -
[\bar q(x)  \imath \gamma_5 q(x)]^2  \right\}.
\label{Hoo}
\end{equation}

\noindent
Since this is a local 4-fermion vertex, the ultraviolet
cut-off must be introduced to regularize the integrals. The
physical interpretation of this cut-off is obvious:
the instanton-induced interaction is operative only when
the squared four-momenta of the quarks   are small. The reason is that
the effective interaction comes from the existence of
the zero modes of quarks (i.e. the zero mass quark is bound
by the instanton  exactly with zero energy). If the three-momentum 
of
the travelling quark is very high but its energy is small, 
it does not see instantons
at all and the instanton-induced interaction between such quarks 
must vanish.
The strength
of the interaction $G$ as well as the ultraviolet cut-off are
directly related to such parameters as an average size of
instantons as well as an average separation between them
\cite{SHURYAK,DP}. This interaction is attractive in $f_0$
and $\pi$ channels (the first and the second terms) and
repulsive in $a_0$ and $\bar \eta$ channels (the third and the
fourth terms). The repulsion in the latter channels must
be contrasted with the attraction in these channels that
is prescribed by perturbative gluon exchange.
 The repulsion in these channels explicitly
breaks the $U(1)_A$.  The interaction
is $SU(2)_L \times SU(2)_R$ symmetric. The first two terms in eq.
(\ref{Hoo}) represent the Nambu and Jona-Lasinio Hamiltonian.
Hence if the interaction is strong enough it can also provide
the spontaneous breaking of chiral symmetry.\\

Summarizing, once the 't Hooft determinant interaction is introduced between
the valence quarks in a meson, it automatically solves the
$U(1)_A$ problem. If this interaction is taken between the
sea quarks in the vacuum, it can provide the spontaneous
breaking of chiral symmetry. For the present context it is 
crucially important that this
interaction is a {\it low-momentum } interaction.
Hence, for the low-lying hadrons, where the typical momenta
of valence quarks are not high, the nonperturbative dynamics due 
to instantons is important.\\

 When we are high
in the spectra,  the three-momenta of valence quarks increase
(contrary to their energy) \footnote{This is natural in the
string picture where practically the whole energy of the hadron
is accumulated in the string while the quarks at the ends have
a large three-momentum.},
hence the 't Hooft interaction between the valence quarks
vanishes and the $ U(1)_A$
symmetry is restored. Similarly, the fast moving valence quarks
do not interact via instantons (or via some other gluonic
interaction that is responsible for chiral symmetry breaking)
with the sea quarks in the vacuum; thus they
decouple from the quark condensates of the vacuum. Consequently the
chiral $SU(2)_L \times SU(2)_R$ symmetry is also 
effectively restored.\\

The QCD vacuum is a very complicated medium and has many
facets (like cubistic paintings). It contains instantons,
other possible topological configurations and mostly the quantum
fluctuations around them. Different probes see different facets
of the vacuum. For example, if one probes the vacuum by  heavy
quarks, those facets that are important for the breaking of
chiral as well as $U(1)_A$
symmetries  become irrelevant. The heavy quarks simply do
not see them. However the aspects of the vacuum that are important
for confinement are relevant in  this case. Indeed, it could be
possible that {\it predominantly} the stochastic structure of 
the QCD vacuum
 \cite{DS} which nicely
 explains the area law of the Wilson loop and also the
 Casimir scaling \cite{B}, does underly the physics of the
 heavy quarks. However, once we probe the vacuum by light quarks,
 in addition other facets (like instantons) become important and
 the physics become reacher. The light quarks do see the instantons
inspite their weight in the full QCD action is very small.\\
 
 As a conclusion, the present results 
 show that high in the spectrum the chiral $SU(2)_L \times SU(2)_R$
symmetry is restored and probably also the $U(1)_A$ one.
It then follows that it is not a confining gluodynamics
in QCD that is responsible for chiral and $U(1)_A$ breakings.

\bigskip

I am grateful to D.V. Bugg for comments on the data \cite{BUGG1,BUGG2,BUGG0}. 
I am also thankful to  C. Lang  for comments on lattice simulations
as well as  to D. Diakonov and T.D. Cohen for valuable
correspondence. The work was supported by the FWF project P14806-TPH
of the Austrian Science Fund.

\section{Appendix}

In this appendix we show how the unmixing of the pure
 $\bar \eta =(u\bar u + d\bar d)/\sqrt{2}$ and
 $\bar \eta_s = s\bar s$ states from the physical
 $\eta$ and $\eta \prime$ mesons is done.
 
 The physical $\eta$ and $\eta \prime$ mesons can be written
 as
 
 $$ \eta =
 (\frac{1}{\sqrt{3}} \cos10 ~ + 
\frac{\sqrt{2}}{\sqrt{3}} \sin10 ~)\bar \eta
 +
 (\frac{1}{\sqrt{3}} \sin10 ~ - 
\frac{\sqrt{2}}{\sqrt{3}} \cos10 ~)\bar \eta_s,$$

$$ \eta \prime =
 (\frac{\sqrt{2}}{\sqrt{3}} \cos10 ~ - 
\frac{1}{\sqrt{3}} \sin10 ~)\bar \eta
 +
 (\frac{\sqrt{2}}{\sqrt{3}} \sin10 ~ + 
\frac{1}{\sqrt{3}} \cos10 ~)\bar \eta_s.$$

Assuming the mass-squared mixing matrix , the masses of physical
$\eta$ and $\eta \prime$ can be found from the eigenvalue problem

$$\left|\begin{array}{cc}
m_{\bar \eta}^2 - m_{ \eta, \eta \prime}^2 & V^2\\
V^2 & m_{\bar \eta_s}^2 - m_{ \eta, \eta \prime}^2
\end{array} \right| =0.
$$

This equation together with the mixing relations above
allow to determine the masses $m_{\bar \eta} \simeq 782$ MeV and
$m_{\bar \eta_s} \simeq 778$ MeV.

\end{document}